# Hardware Architecture of Complex K-best MIMO Decoder

**Mehnaz Rahman**  *mehnaz@tamu.edu*
*Department of ECE*
*Texas A&M University*
*College Station, Tx- 77840, USA*

**Gwan S. Choi**  *gchoi@ece.tamu.edu*
*Department of ECE*
*Texas A&M University*
*College Station, Tx- 77840, USA*

**Abstract**

This paper presents a hardware architecture of complex K-best Multiple Input Multiple Output (MIMO) decoder reducing the complexity of Maximum Likelihood (ML) detector. We develop a novel low-power VLSI design of complex K-best decoder for $8 \times 8$ MIMO and 64 QAM modulation scheme. Use of Schnorr-Euchner (SE) enumeration and a new parameter, Rlimit in the design reduce the complexity of calculating K-best nodes to a certain level with increased performance. The total word length of only 16 bits has been adopted for the hardware design limiting the bit error rate (BER) degradation to 0.3 dB with list size, K and Rlimit equal to 4. The proposed VLSI architecture is modeled in Verilog HDL using Xilinx and synthesized using Synopsys Design Vision in 45 nm CMOS technology. According to the synthesize result, it achieves 1090.8 Mbps throughput with power consumption of 782 mW and latency of 0.044 us. The maximum frequency the design proposed is 181.8 MHz.

## 1.  INTRODUCTION
The introduction of multiple input multiple output (MIMO) is a monumental leap in wireless communication system design. It uses the spatial dimension due to the presence of multiple antenna at the transmitter and receiver ends and provides diversity gain and increased reliability. With the help of MIMO, multiple data can be sent simultaneously through different antennas achieving diversity gain. For a reliable communication, same data can also be sent through multiple antennas. MIMO technology has already been acclaimed by different wireless standards, such as IEEE 802.11n, IEEE 802.16e in order to achieve high data rates. Most of these standards have a specified minimum bit error rate (BER) or packet error rate (PER) to guarantee quality of service (QoS). Such as $10^{-6}$ is specified as maximum tolerable BER according to IEEE 802.11n standard [1].

The main challenge of MIMO system is to design a low-complexity, low-power, high-performance and high-throughput receivers. Several algorithms have been proposed so far to address the issue, offering different tradeoffs between complexity and performance. Among them, maximum-likelihood (ML) detection is the optimum detection method and minimizes the BER through exhaustive search, although its complexity increases exponentially with the number of transmit and receive antennas [2, 3]. On the other hand, linear detectors such as zero-forcing (ZF), the minimum mean squared error (MMSE) have lower complexity with significant performance loss. Hence, a large category of detectors has been proposed trading off between complexity and performance loss, out of which the depth-first and breadth-first search algorithms are well evaluated methods.

The depth-first method like sphere decoder (SD) provides priority to the descent nodes during search process and traces back while reaching to the leaf nodes [4]. On the contrary, the breadth-first scheme such as K-best detector considers a limited number of candidates at each stage in order to proceed to the next stage. So it is a one-pass search with no additional trace backing [5, 6]. Among the two types of search schemes, breadth-first is the most popular in the perspective of implementation due to its constant search complexity.

Recently, lattice reduction (LR) has been proposed in order to achieve high performance as proposed in [7, 8, 9]. LR-aided detector can attain similar diversity as of ML at the cost of performance loss with much less complexity compared to the conventional K-best decoder [10, 11]. Then, it is implemented in complex domain [12]. All of these suboptimal detectors mentioned above are based on hard decision, where data symbols are decided based on the confidence of the detection with no extra information. On the other hand, soft decision schemes calculate the log likelihood ratio (LLR) of each data bit using error correction coding scheme (ECE) and perform the correction. Hence, soft input-soft output (SISO) detectors, suitable for subsequent iterative decoding are introduced in [13]. The method works on the top of the tree search based hard decision and LLR values are calculated using partial available information.

Researchers further improve these SISO detectors with low density parity check (LDPC) decoder [14, 15] in order to reduce the high computing complexity. LLR values for LDPC decoder are first computed from the K best candidates and then, they are fed back to LLR update unit as inputs to the next iteration. This process of iterations is continued until the gain of subsequent iteration becomes saturated. This is called iterative decoding. It can achieve near Shannon performance with less computational complexity compared to other near Shannon decoders [16].

In our previous work [17], a complex K-best iterative MIMO detector was introduced with a new tunable parameter, Rlimit besides list size, K in order to enable adaption of computational complexity with performance gain. For $8 \times 8$ MIMO, it achieves 6.9 to 8.0 dB improvement over real domain K-best decoder and 1.4 to 2.5 dB better performance comparing to conventional complex K-best decoder for 4th iteration and 64 QAM modulation scheme with Rlimit equal to 1 to 4. Furthermore, in [18], a novel study on fixed point realization of iterative LR-aided K best decoder is conducted using MATLAB simulation. The process includes selecting optimized architecture for each sub-module of K-best decoder, and also performing the fixed point conversion to minimize the bit length resulting reduction to hardware cost, power, and area as well. The simulation results show that the total word length of only 16 bits can keep BER degradation within 0.3 dB for MIMO with different modulation schemes.

In this paper a low-power hardware design of iterative complex K-best decoder is presented. The design is specified for 8 × 8 MIMO and 64 QAM modulation scheme with K and Rlimit as 4. The VLSI architecture is modeled in Verilog HDL using Xilinx and synthesized using Synopsys Design Vision in 45 nm CMOS technology. For higher throughput and eliminating dependency, 8 sets of structure for 8 levels are proposed. At the first level, the data is received from antenna and hardware for the other 7 levels fetch the data from the immediate corresponding registers. The proposed architecture design is capable of accomplishing one $8 \times 8$ MIMO signal vector detection every 8 clock cycles. In Synopsys analysis, the design attains the maximum working frequency up to 181.8 GHz and suggests a 1090.8 Mbps data rate with latency as 0.044 us and power consumption of 652 mW for a $8 \times 8$ MIMO system using 64QAM modulation with K and Rlimit equal to 4.

The rest of the paper is organized as follow. In Section II we introduce the algorithm and hardware architecture of complex MIMO decoding algorithm is presented in Section III. Then, Section IV presents the results and Section V concludes this paper with a brief overview.

## 2. SYSTEM MODEL

Let us consider a MIMO system operating in M-QAM modulation scheme and having $N_T$ transmit antenna and $N_R$ receiving antenna as:

$$y = Hs + n, \qquad (1)$$

where $s = [s_1, s_2, \ldots s_{N_T}]^T$ is the transmitted complex vector, $H$ is complex channel matrix and $y = [y_1, y_2, \ldots y_{N_R}]^T$ is $N_R$ dimensional received complex vector [19]. Noise, $n = [n_1, n_2, \ldots n_{N_R}]^T$ is considered as complex additive white Gaussian noise (AWGN) with variance and power $\sigma^2$ and $N_0$ respectively.

The detector solves for the transmitted signal by solving non-deterministic hard problem:

$$\hat{s} = \arg_{\tilde{s} \in S^{N_T}} \min \|y - H\tilde{s}\|^2 . \qquad (2)$$

Here, $\tilde{s}$ is the candidate complex vector, $\hat{s}$ is the estimated transmitted vector [11] and $\|.\|$ denotes 2-norm. This MIMO detection problem can be represented as the closest point problem in [20]. It conducts an exhaustive tree search through all the set of all possible lattice points in $\tilde{s} \in S^{N_T}$ for the global best in terms of partial Euclidean distance (PED) between $y$ and $H\tilde{s}$. Each transmit antenna performs two level of search for real-domain MIMO detection: one for real and the other for imaginary part. However, in complex domain detection method, only one level of search is required for each antenna [17].

ML detector achieves the best performance by attending an exhaustive search through the set of all possible branches from root to node. Hence, its complexity increases exponentially with the number of antennas and constellation bits. Therefore, suboptimal detectors such as LR-aided detector come into consideration.

### 2.1 Lattice Reduction (LR) aided Decoder

Lattice reduction provides more orthogonal basis with short basis vector from a given integer lattice points. Therefore, it reduces the effects of noise and mitigates error propagation in MIMO detection. Eq. (2) is changed to $\hat{s} = \arg_{\tilde{s} \in \mathcal{U}^{N_T}} \min \|y - H\tilde{s}\|^2$ in order to obtain a relaxed search with unconstrained boundary. Here, $\mathcal{U}$ is unconstrained complex constellation set $\{\ldots, -3+j, -1-j, -1+j, 1-j, \ldots\}$. Hence, $\hat{s}$ may not be a valid constellation point. This is resolved by quantizing $\hat{s} = Q(\hat{s})$, where $Q(.)$ is the symbol wise quantizer to the constellation set, $S$.

However, this type of naive lattice reduction (NLD) does not obtain good diversity multiplexing tradeoff (DMT) optimality. Hence, MMSE regularization is employed as proposed in [21, 22], where the channel matrix and received vector are extended as $\bar{H}$ and $\bar{y}$:

$$\bar{H} = \begin{bmatrix} H \\ \sqrt{\frac{N_0}{2\sigma_2^2}} I_{N_T} \end{bmatrix}, \qquad \bar{y} = \begin{bmatrix} y \\ 0_{N_T \times 1} \end{bmatrix}, \qquad (3)$$

where $0_{N_T \times 1}$ is a $N_T \times 1$ zero matrix and $I_{N_T}$ is a $N_T \times N_T$ complex identity matrix [23, 24]. Then, $\hat{s}$ can be represented as:

Mehnaz Rahman & Gwan S. Choi

$$\hat{s} = \arg_{\tilde{s} \in \mathcal{U}^{N_T}} \min \|\bar{y} - \bar{H}\tilde{s}\|^2 . \tag{4}$$

Hence, lattice reduction is applied to $\bar{H}$ to obtain $\widetilde{H} = \bar{H}T$, where T is a unimodular matrix. Eq. (5) then become:

$$\hat{s} = T \arg \min_{\tilde{z} \in \mathcal{U}^{N_T}} \left( \|\tilde{y} - \widetilde{H}\tilde{z}\|^2 + (1+j)_{N_T \times 1} \right), \tag{5}$$

where $\tilde{y} = (\bar{y} - \bar{H}(1+j)_{N_T \times 1})/2$ is the complex received signal vector and $(1+j)_{N_T \times 1}$ is a $N_T \times 1$ complex one matrix. After shifting and scaling, (5) can be represented as $\hat{s} = T\tilde{z} + (1+j)_{N_T \times 1}$. Lattice reduction is considered as NP complete problem. Although polynomial time algorithms such as Lenstra-Lenstra-Lovasz (LLL) algorithm in [25] can find near orthogonal short basis vectors.

### 2.2 Complex K-Best LR-Aided MIMO Detection
Complex K-best LR-aided detection is performed sequentially starting at $N_{th}$-level. First, QR decomposition is applied on $\widetilde{H} = QR$, where Q is a $(N_R + N_T) \times (N_R + N_T)$ orthonormal matrix and R is a $(N_R + N_T) \times N_T$ upper triangular matrix. Then (5) is reformulated as

$$\hat{s} = T \arg \min_{\tilde{z} \in \mathcal{U}^{N_T}} \left( \|\breve{y} - R\tilde{z}\|^2 + (1+j)_{N_T \times 1} \right), \tag{6}$$

where $\breve{y} = Q^T \tilde{y}$. The error at each step is calculated by the PED, which is an accumulated error at a given level of the tree. At each level, K best nodes are selected and passed to the next level for consideration. Finally, the one with minimum PED is chosen out of all the K paths through the tree. The number of valid children for each parent in LR-aided K-best algorithm is infinite. Hence, in our previously proposed algorithm [17], the infinite children issue is addressed using complex on-demand child expansion.

### 2.3 Complex On-demand Expansion
Complex on-demand expansion strategy employs expanding of a node (child) if and only if all of its better siblings have already been expanded and chosen as the partial candidates [8, 26]. It is based on the principle of Schnorr-Euchner (SE) enumerate ion [12, 27]. Hence, K candidates are selected in an order of strict non-decreasing error.

In conventional complex SE enumeration as proposed in [27], expansion of a child can be of two types: Type I, in which the expanded child has same imaginary part as its parent, i.e. enumerating along the real axis; and Type II for all other cases. In our previous work [17], the type of a child is not considered for expansion, instead a new parameter, Rlimit is introduced. The example of improved complex SE enumeration with Rlimit as 3 is given in Fig. 1.

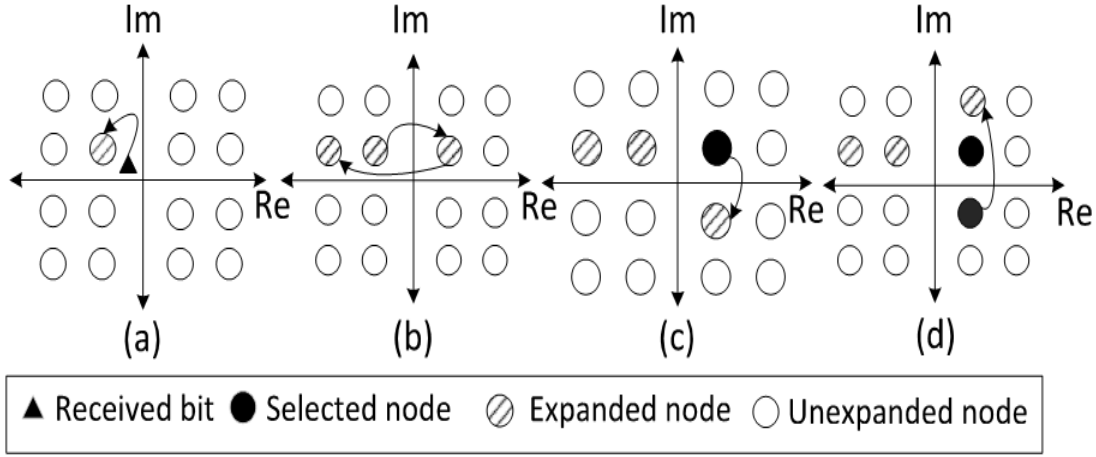

**FIGURE 1**: Improved Complex SE Enumeration with Rlimit as 3.

As shown in Fig. 1, after rounding the received symbol to the nearest integer, real SE enumeration is performed in order to calculate $Rlimit$ candidates. Hence, all the calculated nodes up to $Rlimit$ will have same imaginary value, as demonstrated in Fig. 1(b). Then, the one with minimum PED is selected and expanded only along the imaginary axis using imaginary domain SE enumeration. This process is continued till K nodes are selected at that level of tree, presented in Fig. 1(c)-(d).

The complexity analysis of the improved child expansion proceeds as follows. At any level of tree search, first $KRlimit$ nodes need to be expanded. After that, only imaginary domain SE enumeration will be performed. Hence, considering the worst case, the total number of nodes calculated at each level is $KRlimit + (K - 1)$. For $N_T$ levels, the complexity becomes $N_T K (Rlimit + 1) - N_T$, where for conventional complex decoder and LR-aided real decoder, the complexity is $3N_T K - 2N_T$ and $4N_T K - 2N_T$ respectively [27, 19]. Therefore, introduction of $Rlimit$ offers a re-configurability and tradeoff between complexity and performance.

## 3. ARCHITECTURE PROPOSAL

In this proposed work, a low-power hardware design of iterative complex K-best decoder is presented. The design is specified for 8 × 8 MIMO and 64 QAM modulation scheme with K and Rlimit as 4. For higher throughput and eliminating dependency, 8 sets of structure for 8 levels are proposed. For the first level, the data is received from antenna and hardware for the other 7 levels fetch the data from the immediate corresponding registers.

The proposed architecture for complex K-best decoder consists of two blocks. First one is Data-path block which contains all the processing elements for arithmetic, logical, and sorting operations. The second one is Control-path block which provides synchronization and control signaling. The block diagram of the proposed architecture is shown in Fig. 2.

Mehnaz Rahman & Gwan S. Choi

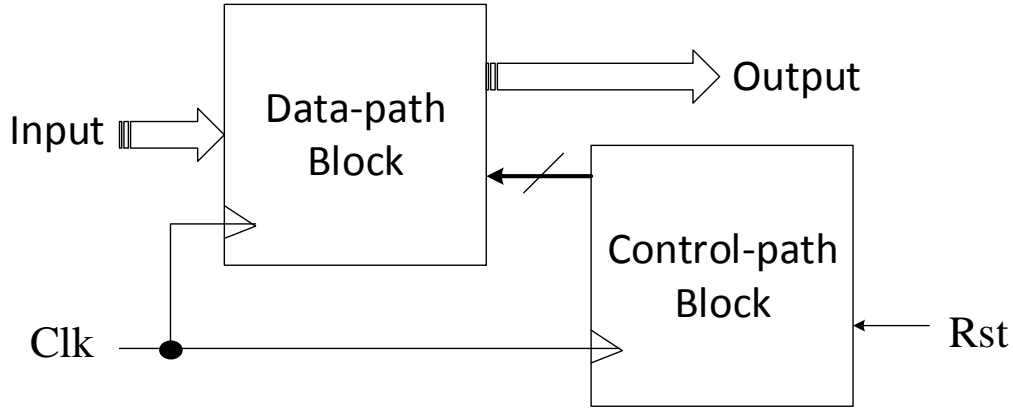

**FIGURE 2:** Block diagram of the proposed architecture.

As presented in Fig. 2, Input includes $\tilde{y}$ and R according to eq. (6) and the Output denotes list and distance representing K-best node list and the cumulative PED distances respectively. Clk is considered as system clock. Additional initialization is done through a reset signal, Rst. The detail explanations of the Data-path block and Control-path block are presented in the subsequent sections.

### 3.1 Data-path Block
The generalized illustration of proposed data-path architecture design is presented below in Fig. 3.

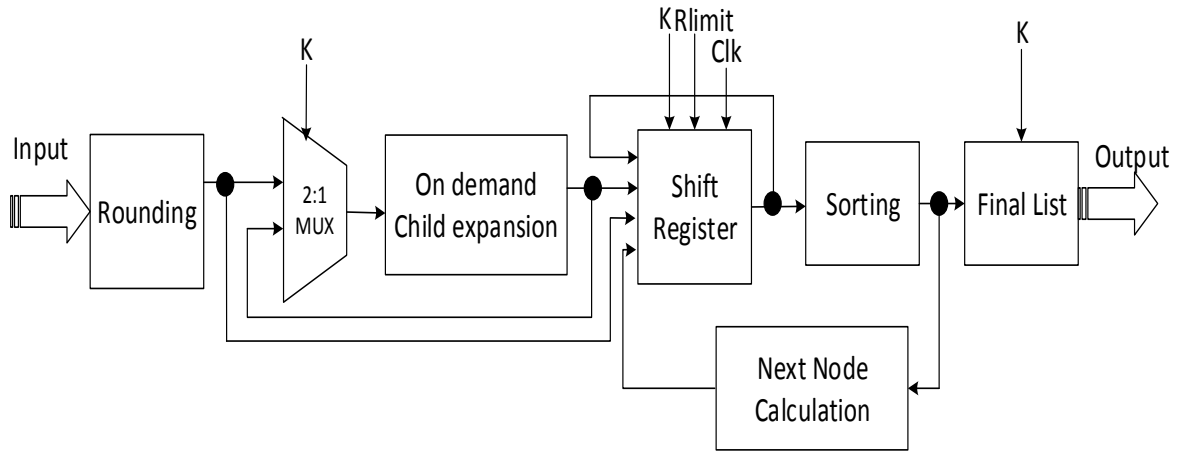

**FIGURE 3:** Block diagram of the data-path architecture.

After receiving the initial input, rounding is first performed as shown in Fig. 3. Then the initial K-best nodes (children) are calculated along the real axis using on demand child expansion. While calculating each node, it is passed to the register. Hence, the register will be initially updated K times with K nodes of real domain. Then, sorting is done to choose the one with minimum distance and selected as a future node for the next level. Hence, the future node is passed to the final list and next child is also calculated from that using on demand expansion along imaginary axis to update the register of the particular index. This sorting and updating the final list as well as register are repeated till K-best nodes are selected for the future candidates of the next level. Therefore, updating the register can be done in four ways: after rounding, after calculating the initial nodes, after calculating the node in imaginary domain, or it can retrieve its previous value.

Figure 3 can be considered as a robust hardware design for all the 8 levels. The generalized illustrations of shift register and sorter are given in Fig. 4(a) and 4(b) respectively.

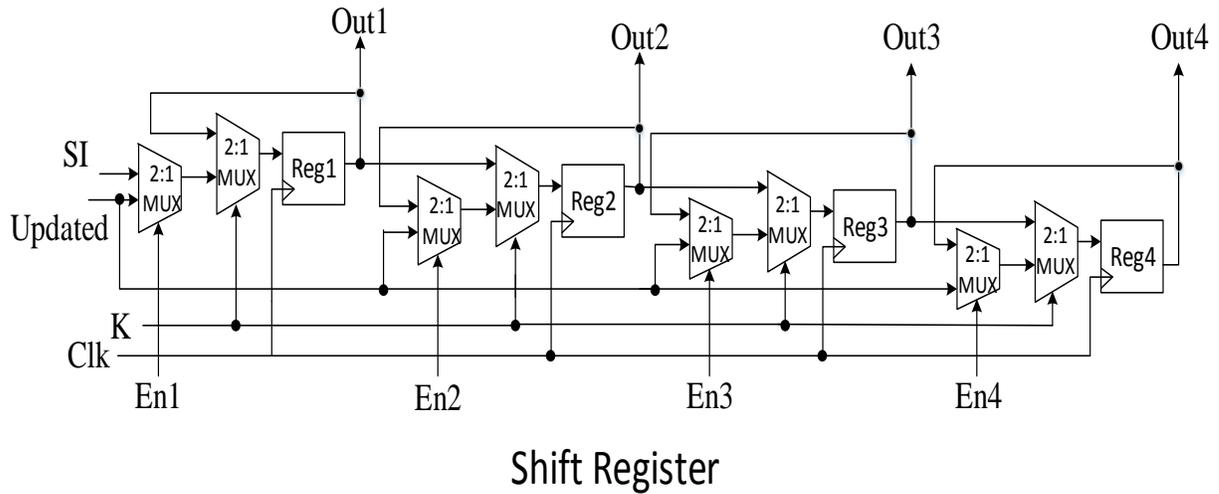

(a)

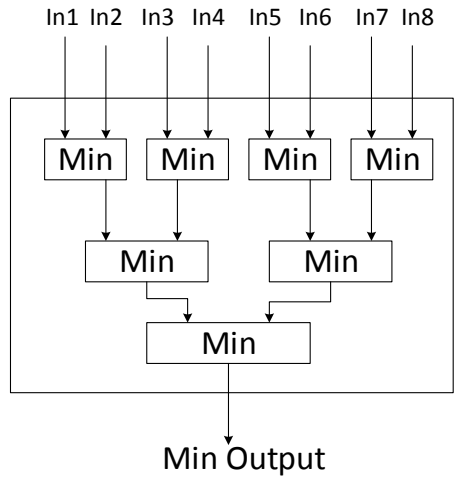

(b)

**FIGURE 4:** Block diagram of the shift register and sorter.

As presented in Fig. 4(a), the operation of shift register is controlled by Updated, K and Rlimit from the control-path block. Initially the shift register will be updated by the serial input (SI) from the on-demand child expansion for K times. Then, after sorting only the corresponding register will be loaded with next best node out of 4 registers. 4 enable signals (En1, En2, En3, and En4) decides which register will be updated. For sorter in Fig. 4(b), a feed-forward pipelinable VLSI architecture is considered for simple implementation. In the proposed work, we also include the pipelining effect among 8 levels of detection to enhance the performance and throughput. The design flow for all the 8 levels of hardware is presented as follows in Fig. 5.

First K best candidates are calculated by level 1 hardware from the input. Then, they are passed to the Reg1. In the meantime, level 1 starts working with new input and level 2 hardware fetches the value from Reg1 and starts performing. This process will go on till the level 8 hardware fetches the value from Reg7 and perform the final output. Detection algorithm proposed in [17] is interdependent and sequential. Hence, 8 units of hardware set need to be used for the 8 antennas to include the pipeline effect and increase the throughput. Since the computational complexity of each level of hardware is low due to elimination of any multiplier and divider [17], use of 8 sets of hardware for 8 antennas does not include high cost in terms of power consumption. The pipelined structure of 8 ×8 MIMO is given in Figure 5.

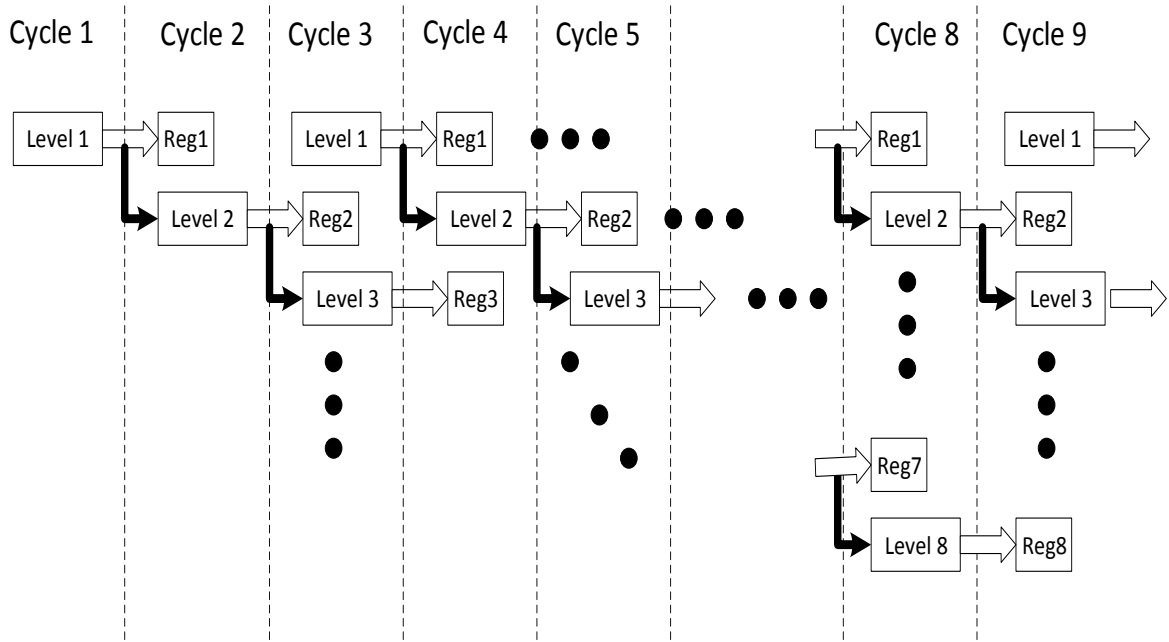

**Figure 5.** Proposed pipelined architecture for 8 ×8 MIMO.

### 3.2 Control-path Block
The Control provides synchronization and control signaling for the data-path block for decoding properly. It consists of a finite state machine that handles all the required control signals for calculating and detecting the K-best nodes at each level. The use of two counter (K, Rlimit) decides the number of node calculation and also required clock cycle. After rounding the initial input, nodes are calculated along the real axis K times and passed to the shift register. Hence, shift register will be updated initially from the real domain child expansion block K times. Then, sorting and next best child calculation along the imaginary domain will be done Rlimit times. Therefore, shift register will again be updated according to the control signal, Rlimit from the control-path block. In this work,

Mehnaz Rahman & Gwan S. Choi

both K and Rlimit are set to 4. Hence, detection at one level requires 8 clock cycles. Finally, the final list size is operated by control signal, K and sent to the next level of detection.

It is worth noting that the proposed architecture is fully pipelined. Hence it can be easily applied in the cases of multicarrier scenarios and each subsequent carrier can be passed to the proposed MIMO detector through pipelining. It can also be applied for different channel conditions with channel estimator if the channel condition is known to the receiver.

## 4. Result

The proposed VLSI architecture is modeled in Verilog HDL using Xilinx and synthesized using Synopsys Design Vision in 45 nm CMOS technology. It is designed for operating with transmitted signal vectors generated from 8 × 8 MIMO and 64 QAM modulation scheme having K and Rlimit set to 4. Simulations for functional validation and verification are done using MATLAB and Xilinx.

As presented in our previous work [18], a novel study on fixed point realization of iterative LR-aided K best decoder is conducted based on simulation. The process includes selecting optimized architecture for each sub-module of K-best decoder, and also performing the fixed point conversion to minimize the bit length resulting reduction to hardware cost, power, and area as well. The simulation results show that the total word length of only 16 bits can keep BER degradation within 0.3 dB for 8 × 8 MIMO and 64 QAM modulation scheme. Hence, in this presented work total word length of 16 bits is considered for the design and implementation approach.

### 4.1 Synthesis Results

The design specs of the proposed complex on-demand K-best decoder attain the requirement of IEEE 802.16e. In each detection, there includes 2 computing stages and complicated operations such as sorting, PED calculation etc. Hence, multiple clock cycles are required for MIMO detection at each level. In our proposed architecture, (K + Rlimit) times, i.e., 8 clock cycles are necessary for detection at every stage. From the synthesis result for 45 nm CMOS technology, 0.044 us latency is observed for detection at each level. Hence, the maximum achievable frequency is 181.8 MHz, leading to 5.5 ns as required time period. Throughput is calculated to be equal to 1090.8 Mbps. If total observed area is divided by the area of a nand gate, the total number of gate count became 63.75 kG.

In order to perform the fair analysis, a normalized hardware efficiency (NHE) is calculated using the following equation [27] and our proposed design achieves NHE as 0.0585.

$$\text{NHE (kG/Mb/ps)} = \frac{\text{core area (kG)}}{\text{scaled throughput (Mb/s)}} \qquad (7)$$

### 4.2 Comparison

The comparison between the proposed complex decoder and the recently proposed MIMO detectors in complex and real domains are tabulated in Table 1.

| Reference | TVLSI 2007 [28] | TCAS 2010 [29] | TVLSI 2010 [30] | JSSC 2010 [31] | JSSC 2011 [32] | TVLSI 2011 [33] | TVLSI 2013 [27] | This work |
|---|---|---|---|---|---|---|---|---|
| Modulation | 16 QAM | 16 QAM | 64 QAM | (4–64) QAM | 64 QAM | 64 QAM | 64 QAM | 64 QAM |
| Antenna | 4 × 4 | 4 × 4 | 4 × 4 | 4 × 4 – 8 × 8 | 4 × 4 | 4 × 4 | 4 × 4 | 8 × 8 |
| Method | K-best | SISO-SD | K-best | MBF-FD (SD) | SISO MMSE-PIC | K-best | Modified K-best | Proposed K-best |
| Domain | Complex | Complex | Real | Complex | Complex | Real | Complex | Complex |
| Process | 0.13 um | 90 nm | 65 nm | 0.13 um | 90 nm | 0.13 um | 0.13 um | 45 nm |
| K | 64 | N/A | 5-64 | N/A | N/A | 10 | 10 | 8* |
| f (max) (MHz) | 270 | 250 | 158 | 198 | 568 | 282 | 417 | 181.8 |
| Throughput (Mb/s) | 100 | 90 | 732 - 100 | 285 - 431 | 757 | 675 | 1000 | 1090.8 |
| Gate count (kG) | 5270 | 96 | 1760 | 350 | 410 | 114 | 340 | 63.75 |
| NHE (kG/Mb/s) | 52.7 | 1.6 | 4.81-35.2 | 1.23-0.81 | 0.78 | 0.17 | 0.34 | 0.0585 |
| Power (mW) | 847 | N/A | 165 | 57-74 | 189.1 | 135 | 1700 | 782 |
| Latency (us) | N/A | N/A | N/A | N/A | N/A | 0.6 | 0.36 | 0.044 |
| Hard/ soft | Soft | Soft | Hard | Soft | Soft | Hard | Hard | Hard |

*In our proposed design, both K and Rlimit are equal to 4.

**TABLE 1:** Design Comparison of the proposed design with previous works.

The table 1 shows that our proposed architecture requires less power and lower latency with higher throughput comparing with all other previous works for both real and soft domain. If we consider [27] as the most updated VLSI architecture published so far in complex domain, the proposed architecture outstands the performance in terms of gate count, power consumption and so on. The architecture implemented in [27] is for 4 x 4 MIMO with 64 QAM modulation scheme and K as 10 using 0.13 um technology, where our proposed design is for 8 x 8 MIMO with 64 QAM modulation and list size of 8 and synthesized using 45 nm technology. Hence, even with less size, the proposed one can achieve higher throughput which 1090.8 Mbps compared to that of [27]. The power consumption is 2.17x less with the requirement of 1/5[th] of the gates compared to the one in [27]. The gate count of proposed decoder is 63.75 kG, where in [27] it is equal to 340 kG. The latency is 8.1x less compared to the published one; although the maximum achievable frequency for our proposed decoder is 181.8 MHz, where the architecture in [27] can attain higher frequency which is 417 MHz. As shown in [16], we can improve the performance by increasing the value of K and

Mehnaz Rahman & Gwan S. Choi

Rlimit with allowing more complexity. The complexity can also be reduced with less K and Rlimit on the contrary trading off some performance loss. Moreover, the proposed architecture provides more re-configurability in terms of complexity and performance.

## 5. CONCLUSION

In this paper, a VLSI architecture of complex domain K-best decoder is proposed exploiting the improved complex on-demand child expansion. It includes an additional parameter, Rlimit in order to trade-off the complexity of computation with improvement in BER performance. Although the proposed approach is scalable to any MIMO configuration and constellation order, the design is specified for 8 × 8 MIMO with 64 QAM modulation scheme for K and Rlimit equal to 4.

The proposed VLSI architecture is modeled in Verilog HDL using Xilinx and synthesized using Synopsys Design Vision in 45 nm CMOS technology. Simulations for functional validation and verification are done using MATLAB and Xilinx. Next, the comparison of the performance with the previous works is mentioned for fair evaluation. Our design approach achieves 1090.8 Mbps throughput with power consumption of 782 mW and latency if 0.044 us. The number of gate count required is 63.75 kG and it can achieve frequency up to 181.8 MHz.

Future work of this proposed architecture includes evaluating the detector performance and synthesis result with improved and modified design for each critical block (such as sorter, PED calculation etc.).